\documentclass[referee]{aa}
\usepackage{natbib}
\usepackage{amssymb}
\usepackage{graphicx}
\newcommand{\micron}{$\mu$m}
\title{$AKARI$ detections of hot dust in luminous infrared galaxies}
\subtitle{Search for dusty active galactic nuclei}
\author{S. Oyabu\inst{1}\thanks{E-mail: oyabu@u.phys.nagoya-u.ac.jp}, D. Ishihara\inst{1}, M. Malkan\inst{2}, H. Matsuhara\inst{3}, T. Wada\inst{3},
  T. Nakagawa\inst{3}, Y. Ohyama\inst{4}, Y. Toba\inst{3,5},
  T. Onaka\inst{6}, S. Takita\inst{3,5}, H. Kataza\inst{3}, I. Yamamura\inst{3},
  and M. Shirahata\inst{3}} 
\institute{
  Graduate School of Science, Nagoya University, Furo-cho, Chikusa-ku,
  Nagoya, Aichi 464-8602 Japan \and
 Department of Physics and Astronomy, University of California, Los
 Angeles, CA 90095-1547, USA \and
  Institute of Space and Astronautical Science, Japan
  Aerospace Exploration Agency, 3-1-1 Yosinodai, Chuo-ku, Sagamihara, Kanagawa
  252-5210, Japan 
\and Academia Sinica, Institute of
  Astronomy and 
  Astrophysics, Taiwan \and Department of Space and Astronautical
  Science, the Graduate University for Advanced Studies (Sokendai),
  3-1-1 Yoshinodai, Chuo-ku, Sagamihara, Kanagawa 252-5210, Japan \and
  Department of Astronomy, 
  University of Tokyo, 7-3-1 Hongo, Bunkyo-ku, Tokyo, 113-0033, Japan
  }
\date{Received / Accepted }

\abstract{}{ We present a new sample of
  active galactic nuclei (AGNs) identified 
  using the catalog of the $AKARI$
 Mid-infrared(MIR) All-Sky Survey. Our MIR search
 has the advantage of detecting AGNs that are obscured at optical
 wavelengths by extinction.} 
{We first selected $AKARI$
  9\micron~excess sources with 
  $F(9\mu\mathrm{m})/F(K_S)>2$ where $K_S$ magnitudes were taken from the Two Micron
  All Sky Survey. We then obtained follow-up  near-infrared spectroscopy with the
$AKARI$/IRC to confirm that the excess is
caused by hot dust. We also
obtained optical spectroscopy with the Kast Double
Spectrograph on the Shane 3-m telescope at Lick Observatory.}
{On the basis of these
observations, we detected hot dust with a characteristic temperature
of $\gtrsim 500\mathrm{K}$ in 
two luminous infrared galaxies.
The hot dust is suspected to be associated with AGNs that exhibit
their nonstellar activity not in the optical, but in the near- and mid-infrared
bands, i.e., they harbor buried AGNs. 
The host galaxy stellar masses of $\sim 4-6 \times 10^9\ \mathrm{M_{\odot}}$ are
small compared with the hosts in optically-selected AGN populations.
These objects were missed by previous surveys, 
demonstrating the
power of the $AKARI$ MIR All-Sky Survey to widen AGN searches 
to include more heavily obscured objects.
The existence of
multiple dusty star clusters with massive stars cannot be completely ruled out with our
current data.}{}
\titlerunning{$AKARI$ detections of hot dust in galaxies}
\authorrunning{S. Oyabu}
\begin{document}
\maketitle

\section{Introduction}

Many observations have found evidence of the presence of a large number of heavily
obscured active galactic nuclei (AGNs). It is
found that a significant number of ultra-luminous infrared galaxies 
contain AGNs at their centers \citep[e.g.][]{sanders96, lutz98, imanishi08}. 
Various hard X-ray and soft gamma-ray observations
\citep{maiolino98,risaliti99,malizia09} indicate that about 80 percent
of the AGNs in the local Universe are obscured. AGN synthesis
models of the X-ray background postulate their existence  in order to
explain the flat spectrum of the hard X-ray background \citep[e.g.][]{ueda03}. 
These black holes contribute to the local
black hole mass density \citep{fabian99}, and
 are potentially important contributors to the growth of 
supermassive black holes throughout the history of the universe.
However, 
the nature of this population, even in the local universe, is only
poorly understood, because of the strong selection bias against finding
them at optical wavelengths. 

Mid-infrared(MIR) AGN searches can
overcome this obstacle by penetrating through dust extinction to
identify most of the AGN population, including Type 2 Seyferts and buried AGNs.
The original IRAS 12\micron~active galaxy samples \citep{spinoglio89,
  rush93} provide an unbiased sample of local active
galaxies. Using the $ISOCAM\ parallel\ mode\
survey$ of 10 square degrees at 6.7\micron(LW2), 
\citet{leipski05,leipski07} succeeded in finding redder Type 1 AGNs
as well as Type 2's. Several searches have also
been performed using the near- and mid-infrared bands in the Spitzer Space
Telescope 
\citep[e.g.][]{lacy04,alonso06,polletta06}. 

$AKARI$ performed an all-sky survey at 9 and 18\micron~as well as
at four far-infrared (FIR) bands (65, 90, 140, and 160\micron). It provided
improvements of about one order of magnitude compared to that of IRAS
in both spatial resolution and
sensitivity in the mid-infrared bands. The details of the survey are
described in \citet{ishihara09}.

In this paper, we present the first results of our search for AGNs
based on this $AKARI$ MIR All-Sky Survey. We discovered two galaxies LEDA
84274 and IRAS 01250+2832, which have a compact hot $\gtrsim$500 K
dust component. The hot dust
component may be heated by the central engine of the AGN,
even though their optical spectra do not show any AGN characteristics. 

The observations, data reduction, and results are
described in Section \ref{sec:obs}.
In Section \ref{sec:multi}, we describe the multiwavelength properties of
the two galaxies and in Section \ref{sec:hot} the hot dust components we found in the
galaxies are discussed. 
The discussion is presented in Section
\ref{sec:dis}, and a summary is given in Section \ref{sec:sum}.
Throughout the paper, we assume a flat cosmology with $\Omega = 0.3$,
$\Lambda=0.7$, and $H_0 = 70\ \mathrm{km}\ \mathrm{s}^{-1}\
\mathrm{Mpc}^{-1}$. 

\section{Observations and results}
\label{sec:obs}

The initial identification of the $AKARI$ MIR All-Sky Survey sources
involves association with the Two Micron All Sky Survey (2MASS) catalog \citep{skrutskie06}. This
search highlights unusually red $AKARI$ MIR sources with $F(9\mu\mathrm{m})/F(Ks)
> 2$,  at high Galactic latitudes,
$|b|>30^{\circ}$ after excluding regions around the Large and Small Magellanic
Clouds. 
To examine the origin of the excess of $F(9\mu\mathrm{m})/F(Ks) > 2$, we
performed follow-up observations with the $AKARI$ near-infrared(NIR)
spectrometer. 

Using the $AKARI$ spectra
of the 2.5-5 \micron~wavelength range, 
AGNs can be distinguished by their red continuum emission, while
strong polycyclic aromatic hydrocarbons (PAH) emission 
is detected in star-forming galaxies.
To measure the redshift and search optically for AGN or star-formation signatures, 
optical spectra were also taken with the Share 3m telescope at
the Lick Observatory.

During these follow-up observations, we discovered two buried AGN
candidates that 
have steep red NIR continuum from hot dust, but do not show 
any AGN features in optical spectra, such as strong high-ionization
emission lines.
Table \ref{tab:pro} summarizes the properties of our targets, including
 the fluxes in the $AKARI$ All-Sky MIR and FIR Survey
catalogs \citep{ishihara09,yamamura09}. LEDA 
84274 is identified as the FIR source, IRAS 14416+6618. 
Previous optical spectroscopy of LEDA 
84274 \citep{kim95,vielleux95} reported a
low-ionization spectrum with strong Balmer lines, indicating an
HII-region-like or star-forming galaxy at z=0.0377.
Thus, the gas emitting these optical lines is
photoionized by early-type stars. 
For IRAS 01250+2832, this is its first reported identification as
a galaxy at $z=0.043$. 

\begin{table}
    \caption{Data for LEDA 84274 and IRAS 01250+2832}
    \label{tab:pro}
    \begin{tabular}{lll}
        \hline
        \hline
        Property & \object{LEDA 84274} & \object{IRAS 01250+2832} \\
        \hline
        R.A.(J2000)$^a$ & 14 42 34.88 & 01 27 53.95  \\
        Dec.(J2000)$^a$ & +66 06 04.3 & +28 47 51.0 \\
        Redshift        & 0.0377      & 0.043\\
        $J$(mag)$^b$    & 13.756 $\pm$ 0.054  & 14.774 $\pm$ 0.117 \\
        $H$(mag)$^b$    & 13.293 $\pm$ 0.090  & 13.941 $\pm$ 0.131 \\
        $Ks$(mag)$^b$   & 12.744 $\pm$ 0.083  & 13.462 $\pm$ 0.169 \\
        IRAS 12\micron (Jy)$^c$  & $0.10\pm 0.03 $ (2)  & $<$ 0.11 (1)\\
        IRAS 25\micron (Jy)$^c$  & $0.56\pm 0.03$(3)  & $0.28 \pm 0.05$(2) \\
        IRAS 60\micron (Jy)$^c$  & $2.19\pm 0.20$(3)  & $0.52 \pm 0.03$(3) \\
        IRAS 100\micron (Jy)$^c$ & $1.80 \pm 0.14$(2)  & $<$ 0.76 (1) \\
        $AKARI$ 9\micron (Jy)$^d$  & $0.084\pm 0.007$ & $0.105\pm 0.010$ \\
        $AKARI$ 18\micron (Jy)$^d$ & $0.344\pm 0.016$ & $0.182\pm 0.017$ \\
        $AKARI$ 65\micron  (Jy)$^e$ & $1.89 \pm 0.22$(1) & $< 3.2 $$^i$ \\
        $AKARI$ 90\micron  (Jy)$^e$ & $1.51 \pm 0.04$(3) & $< 0.55$$^i$ \\
        $AKARI$ 140\micron (Jy)$^e$ & $1.79 \pm 0.26$(1) & $< 3.8 $$^i$ \\
        $AKARI$ 160\micron (Jy)$^e$ & $< 7.5$$^i$ & $<7.5$$^i$ \\
        $L_{\mathrm{IR}}(\mathrm{L}_{\sun})$$^f$ & $1.8\times
        10^{11}$ & $1.0 \times 10^{11}$ \\
        $L_{\mathrm{IR}}(\mathrm{L}_{\sun})$$^g$ & $1.8\times
        10^{11}$ & $0.9 \times 10^{11}$ \\
        $D_{\mathrm{n}}(4000)$$^h$ & $1.10\pm0.02$ & $1.63\pm0.08$ \\
        \hline
    \end{tabular}
    \begin{itemize}
    \item[$^a$] The 2MASS coordinates. Units of right ascension are
        hours, minutes, and seconds, and units of declination are
        degrees, arcminutes, and arcseconds. 
    \item[$^b$] The 2MASS magnitudes in All-Sky Extended Source
        Catalog
    \item[$^c$] Fluxes are from the IRAS Faint
        Source\citep{moshir92}. Numbers in parentheses indicate
        the flux qualities. 
    \item[$^d$] $AKARI$/IRC Mid-infrared All-Sky Survey Point Source
        Catalog Ver. 1 \citep{ishihara09}
    \item[$^e$] $AKARI$/FIS FIR All-Sky Survey Bright Source
        Catalog Ver. 1 \citep{yamamura09}. The number in parentheses indicates
        the flux qualities. (3) denotes a clear detection, and (1) indicates the
        flux at the position. 
    \item[$^f$] The 8-1000 \micron~infrared luminosity
        $L_{\mathrm{IR}}=4\pi D_{L}^2 F_\mathrm{IR}$ is computed
        using IRAS fluxes,
        $F_{\mathrm{IR}}=(1.8\times10^{-14})(13.56\times
        F_{\mathrm{12}}+5.26\times F_{\mathrm{25}}+2.54\times
        F_{\mathrm{60}}+1.0\times F_{\mathrm{100}})\ \mathrm{W}\
        \mathrm{m}^{-2}$\citep{kim95}. We used IRAS upper
        limits when IRAS did not detect a source.  
    \item[$^g$]
 The 8-1000 \micron~infrared luminosity is derived
        from the blackbody fits. See Sections \ref{sec:leda}
        and \ref{sec:iras} for details.
    \item[$^h$] $D_{\mathrm{n}}(4000)$ is the discontinuity of the
        spectrum around 
        4000\AA. See Section \ref{sec:lick} for details.
    \item[$^i$] Detection limits in \citet{yamamura10}, which are
        estimated from the peaks of $\log N -\log S$ plots,
        corresponding to  $\sim 90$ percent completeness.
    \end{itemize}
\end{table}

\subsection{$AKARI$ near-infrared spectroscopy}

We performed NIR spectroscopy using the InfraRed Camera
(IRC) \citep{onaka07,ohyama07} onboard the $AKARI$ satellite 
\citep{murakami07}. 
We used the IRC channel NIR, which uses a 512 $\times$ 412 InSb array
and the astronomical observation template Z4 (AOT Z4) designed
for spectroscopy. AOT Z4 replaces the imaging filters by the
transmission-type 
dispersers on the filter wheel to take NIR spectra. 
The NIR grism (NG) was set to cover the wavelengths
of 2.5 - 5\micron~at a resolution of about 100. The
1\arcmin~$\times$ 1\arcmin~slit was used to avoid any overlap with
other sources.

$AKARI$ NIR spectra were obtained during December 2008 - July
2009. The total exposure times for both objects are 792 sec.
We observed both objects twice in
the $AKARI$ warm (Phase3) mission period to correct hot pixels
caused by the relatively high temperature ($\sim$ 40K) of the detector. 

The data were processed using the IRC Phase3 dedicated data reduction
package, IRC\_SPECRED Ver. 20090211 \citep{ohyama07}. Dark subtraction, 
linearity correction, flatfield correction, and various image anomaly corrections 
were first performed. Multiple exposures were then coadded. After performing 
wavelength- and flux-calibration, the spectra of the object were extracted.
The aperture size for the spectrum extraction was set to be
3 pixels (4.5\arcsec).
Aperture corrections were also applied at the end of the processing. 
The present data reduction gives an uncertainty in wavelength of $\sim
0.01$\micron~ and the error in the absolute flux is smaller than 20 percent.

\begin{figure*}
    \includegraphics[width=17cm]{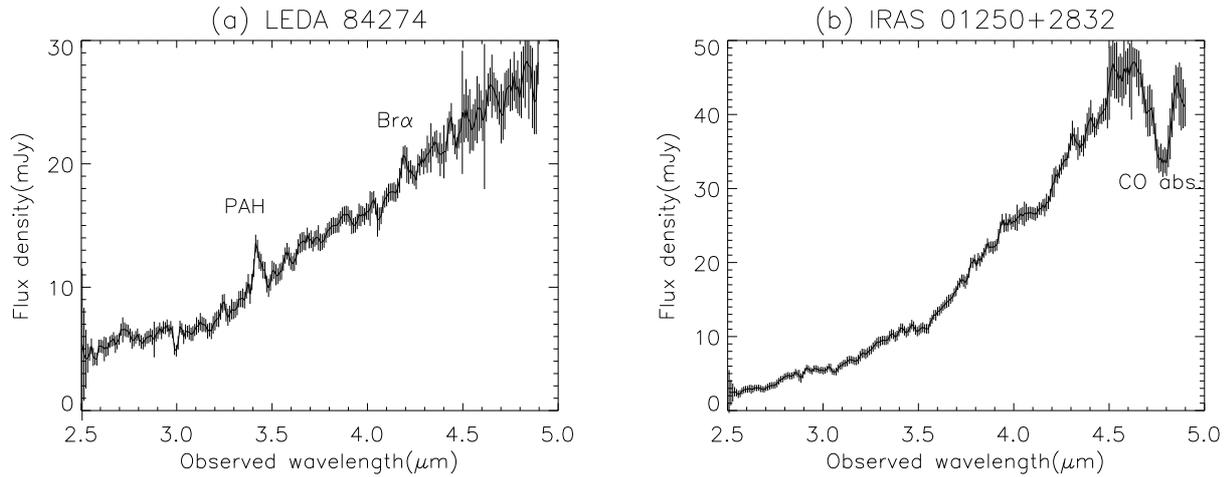}
    \caption{$AKARI$ NIR spectra of (a) LEDA 84274 and (b) IRAS
      01250+2832 in the observed wavelength frame. 
      The thin vertical lines represent 1$\sigma$ errors.} 
    \label{fig:nir_spe}
\end{figure*}

Figure \ref{fig:nir_spe} shows the calibrated spectra of LEDA 84274
and IRAS 01250+2832. Both objects display steep red continuum. We find
emission lines of PAH at 3.3\micron~and
Br$\alpha$ at 4.05\micron~for LEDA 84274. The measured line fluxes
are given in Table 
\ref{tab:akari_leda}. For IRAS 01250+2832, the R-branch of CO ro-vibrational
absorption is detected at 4.75\micron. Upper limits to the fluxes of
the PAH bands at 3.3\micron~and Br$\alpha$ at 4.05\micron~are given in
Table 2. 


\begin{figure*}
    \includegraphics[width=17cm]{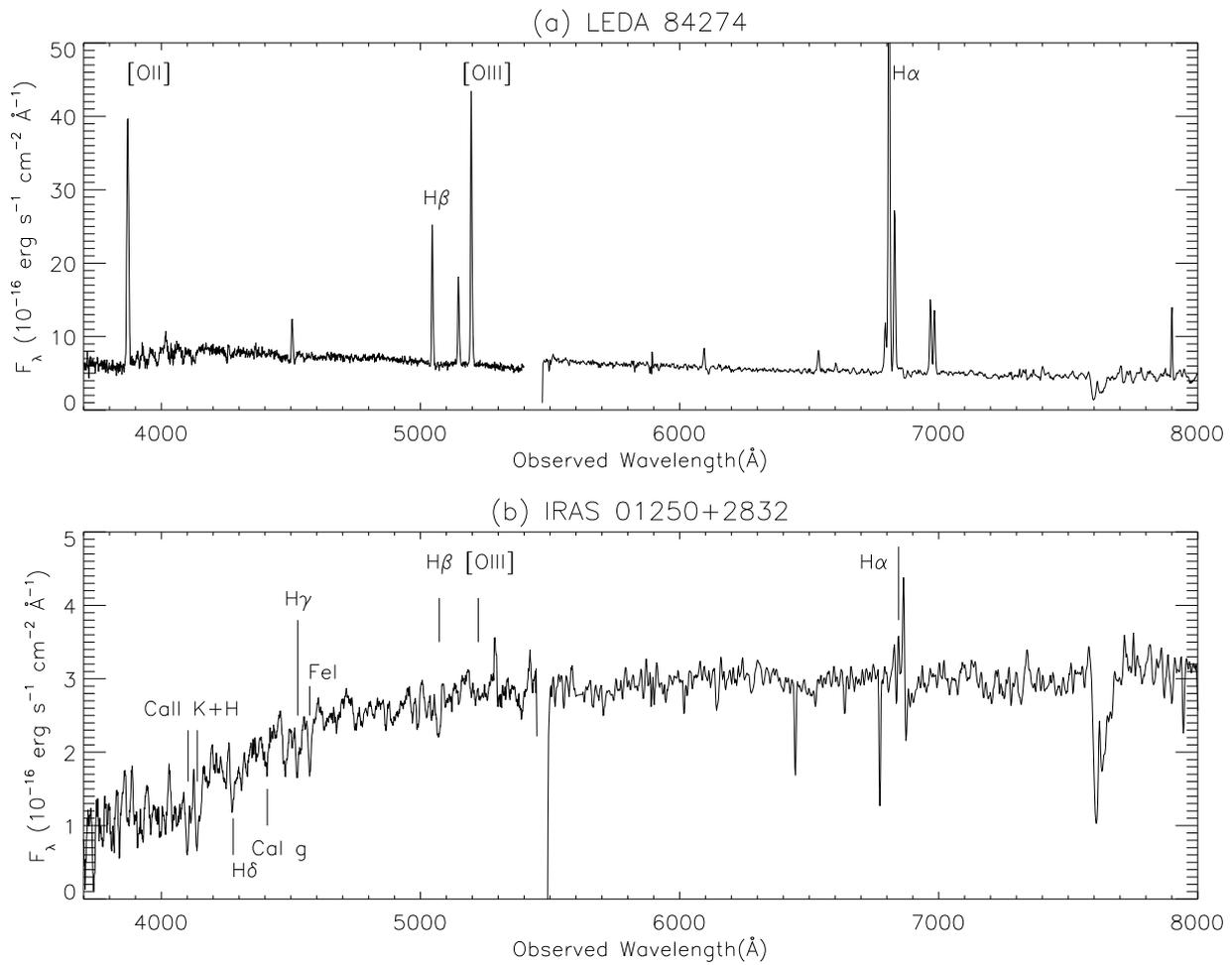}     
    \caption{Optical spectra of (a) LEDA 84274 and (b) IRAS
      01250+2832. In (b), the positions of the
      H$\beta$4861\AA~and [O III] 5007\AA~lines are indicated. }
    \label{fig:opt_spec}
\end{figure*}

\begin{table*}
    \centering
    \caption{$AKARI$ NIR emission features of LEDA 84274 and IRAS 01250+2832}
    \label{tab:akari_leda}
    \centering
    \begin{tabular}{crrr}
        \hline\hline
        Name & Obs. Wave.$^a$ & Flux$^b$ & Obs. EW$^c$ \\
        & (\micron)           &  ($10^{-16}\ \mathrm{ergs}\
        \mathrm{s}^{-1}\ \mathrm{cm}^{-2}$)   & (\micron) \\
        \hline
        \multicolumn{4}{c}{LEDA 84274}\\
        \hline
        PAH 3.3\micron & 3.416$\pm$0.010 & 437 $\pm$  30 & 0.018 $\pm$
        0.001\\
        Br$\alpha$ 4.05\micron    & 4.200$\pm$0.010 & 178 $\pm$ 26  &
        0.060 $\pm$ 0.009\\
        \hline
        \multicolumn{4}{c}{IRAS 01250+2832}\\
        \hline
        PAH 3.3\micron & ... &  $<75$$^d$ & $<0.003$$^d$ \\
        Br$\alpha$ 4.05\micron & ... & $<100$$^d$& $<0.001$$^d$ \\        
        \hline
    \end{tabular}
    \begin{itemize}
    \item[$^a$] Observed wavelength.
    \item[$^b$] Line flux and 1$\sigma$ error in the measurement. 
    \item[$^c$] Observed equivalent width in \micron.
    \item[$^d$] Upper limits(3 $\sigma$) are estimated with the line
        widths for LEDA 84274. 
    \end{itemize}
\end{table*}

\subsection{Lick optical spectroscopy}
\label{sec:lick}

We performed optical spectroscopy
of LEDA 84274 and IRAS 
01250+2832 using the Kast Double Spectrograph  
on the Shane 3-m telescope at the Lick Observatory in 2009 March and
August, respectively. This spectrograph has two separate parallel
channels - one 
optimized for the blue and the other for the red.
The 600/4310 grism and the 600/7500 grating were used in the blue and red
spectrometers, respectively. The wavelength coverage 
was 3700  - 8300 \AA~with a small gap at around 5500
\AA~produced by the dichroic beamsplitter.
The exposure times for both galaxies were 1500 sec.

The spectra of both LEDA 84274 and IRAS 01250+2832 are shown in Figure
\ref{fig:opt_spec}. We identify the strong emission lines in LEDA
84274 and IRAS 01250+2832 as shown in Table \ref{tab:opt_leda}. On the
basis of these identifications, we determine the redshift of 
z=0.0377 for LEDA 84274, which is consistent with the redshift previously determined by
\citet{kim95}. 
The line ratios of the present results are consistent
with previous ones,
while the emission line flux ratios relative to those of the H$\alpha$ line in
bluer part of the spectra are slightly larger than those in \citet{kim95}.  
The spectrum of IRAS 01250+2832 shows a series of
absorption lines in the blue part of its spectrum. We identify these absorption lines as
CaII H+K, H$\delta$, 
CaI g-band, H$\gamma$, and Fe I as shown in Figure
\ref{fig:opt_spec}(b). Although H$\alpha$ 6563\AA~and [N II]
6548,6583\AA~emission lines are detected,
no other 
emission lines are detected.  We use the absorption 
lines to estimate the redshift of $z=0.043$. 

To examine the galaxy stellar population, we measure the 4000\AA~discontinuity.
This discontinuity is originally defined by 
\begin{equation}
    \label{eq:d4000}
    D(4000)=\frac{(\lambda_2^--\lambda_1^-)\int_{\lambda_1^+}^{\lambda_2^+}F\nu
    d\lambda}{(\lambda_2^+-\lambda_1^+)\int_{\lambda_1^-}^{\lambda_2^-}F\nu
    d\lambda},
\end{equation}
where
$(\lambda_1^-,\lambda_2^-,\lambda_1^+,\lambda_2^+)=(3750,3950,4050,4250)$
\AA
\citep{bruzual83}. A definition using narrower continuum bands
$(\lambda_1^-,\lambda_2^-,\lambda_1^+,\lambda_2^+)=(3850,3950,4000,4100)$
was introduced by \citet{balogh99} and is widely used. The
advantage of the 
narrower bands is that the index is less sensitive to  reddening.
We adopt the narrow-band definition, and denote this index as
$D_{\mathrm{n}}(4000)$. 
The resultant $D_{\mathrm{n}}(4000)$ indices are 1.10$\pm$0.02 and 1.63$\pm$0.08 for
LEDA 84274 and IRAS 01250+2832, respectively. 
On the basis of these indices, LEDA 84274 has the spectrum of a late-type galaxy,
and IRAS 01250+2832 that of an elliptical galaxy.

\begin{table*}
    \caption{Optical emission lines of LEDA 84274 and IRAS 01250+2832}
    \label{tab:opt_leda}
    \centering
    \begin{tabular}{crrrrr}
        \hline\hline
        Name & Obs. Wave.$^a$ & Flux & Obs. EW$^b$ & ratio with H$\alpha$$^c$ &
        ratio with H$\alpha$$^d$ \\
             & (\AA)           &  ($10^{-16}\ \mathrm{ergs}\
             \mathrm{s}^{-1}\ \mathrm{cm}^{-2}$)   & (\AA) & This
             work & \citet{kim95}\\
        \hline
        \multicolumn{6}{c}{LEDA 84274}\\
        \hline
        $[$OII$]$       & 3870 & 317.0$\pm$ 1.9 &   54.6$\pm$0.3 &
        0.58 & 0.30 \\
        H$\gamma$       & 4504 &  46.5$\pm$ 1.4 &    7.4$\pm$0.2 &
        0.08 & ...  \\
        H$\beta$        & 5045 & 120.5$\pm$ 1.1 &   20.6$\pm$0.2 &
        0.22 & 0.18 \\
        $[$OIII$]$      & 5146 &  81.9$\pm$ 1.1 &   13.5$\pm$0.2 &
        0.15 & 0.12 \\
        $[$OIII$]$      & 5196 & 240.0$\pm$ 1.1 &   40.6$\pm$0.2 &
        0.44 & 0.37 \\
        HeI      & 6094 &  18.8$\pm$ 0.5 &  3.2$\pm$0.1 & 0.03 & ...\\
        $[$OI$]$     & 6536 &  23.0$\pm$ 0.4 & 4.4$\pm$0.1 & 0.04 & 0.04 \\
        $[$OI$]$     & 6603 &   8.5$\pm$ 0.4 &    1.6$\pm$0.1 & 0.02 & ...\\
        $[$NII$]$    & 6794 &  61.1$\pm$ 0.6 &   11.8$\pm$0.1 & 0.11 & 0.10 \\
        H$\alpha$   & 6808 & 549.0$\pm$ 0.5  &  106.1$\pm$0.1 & 1.00  & 1.00 \\
        $[$NII$]$    & 6830 & 178.0$\pm$ 0.5 &   34.6$\pm$0.1 & 0.32 & 0.31 \\
        $[$SII$]$    & 6968 &  78.9$\pm$ 0.4 &   15.5$\pm$0.1 & 0.14 & 0.14 \\
        $[$SII$]$    & 6983 &  65.1$\pm$ 0.4 &   12.8$\pm$0.1 & 0.12 & 0.11 \\
        \hline
         \multicolumn{6}{c}{IRAS 01250+2832}\\
        \hline
        $[$NII$]$    & 6828 &   4.3$\pm$ 0.8 &   1.5$\pm$0.3 & 1.02 & ... \\
        H$\alpha$    & 6844 &   4.2$\pm$ 0.8 &   1.4$\pm$0.3 & 1.00 & ... \\
        $[$NII$]$    & 6864 &  12.1$\pm$ 0.8 &   4.2$\pm$0.3 & 2.88 & ... \\
        \hline
       \\
    \end{tabular}
    \begin{itemize}
        \item[$^a$]Observed wavelength. A typical error of
            measurements is less than 1\AA.
        \item[$^c$]Observed equivalent width. 
        \item[$^c$]Emission line ratio with H$\alpha$ in this work. A
            typical measurement error is less than 0.01.
        \item[$^c$]Emission line ratio with H$\alpha$ in \citet{kim95}. 
    \end{itemize}
\end{table*}

\section{Multiwavelength properties}
\label{sec:multi}

The $AKARI$ NIR spectroscopic observations display a steep red
continuum  in both galaxies, while
they have different characteristics in the optical and 
FIR.  The following subsections describe the
properties of each galaxy. 

\subsection{LEDA 84274}
\label{sec:leda}

The various properties of LEDA 84274 indicate that it currently has a high star-formation rate. 
The optical emission line ratios such as [NII]6583\AA/H$\alpha$
vs. [OIII]5007\AA/H$\beta$, [SII]6716\AA+6731\AA/H$\alpha$ vs.
[OIII]5007\AA/H$\beta$, and [OI]6300\AA/H$\alpha$
vs. [OIII]5007\AA/H$\beta$ are in the range of a star-forming galaxy
according to 
\citet{vielleux95}, who used the spectrum taken by
\citet{kim95}. 
The PAH emission band detected at 3.416 \micron~is known to
be associated with star-formation activity. FIR detections
with IRAS and $AKARI$ are also consistent with a high star-formation rate.
The 20cm flux density in the NRAO VLA Sky
Survey \citep[NVSS;][]{condon98} is reported to be $4.2 \pm 0.4$mJy. 
The logarithmic ratio
of the FIR to radio continuum flux densities $\log
{[F_{fir}/(3.75\times10^{12}\mathrm{Hz})/f_{\nu}(20\mathrm{cm})]}=2.7
\pm 0.1$, where $F_{fir}$ is the flux between 42.5 and
122.5\micron\footnote{$F_{fir}=1.26\times10^{-14}\times[2.58f_{\nu}(60\mu\mathrm{m})+f_{\nu}(100\mu\mathrm{m})]$
where $f_{\nu}$ are the flux densities in Jy, and $F_{fir}$ is in $\mathrm{W}\mathrm{m}^{-2}$.}\citep{helou85}, is 
consistent with 
$2.4\pm 0.2$ of the star-formation activity
in normal, infrared, and luminous infrared galaxies including
ultra-luminous infrared galaxies \citep{sanders96}. 
This relative weakness of radio
emission in LEDA 84274 is clearly inconsistent with the ratio (2.1) 
observed in local Seyfert galaxies \citep{rush96}.

Before we estimate the star-formation rate in LEDA 84274, we estimate the
amount of reddening in the emission region, 
based on the Balmer line ratio, H$\alpha$/H$\beta$.
The equivalent widths of the Balmer lines are so large that we did
not apply any correction for the underlying stellar absorption lines.
The result is
$A_V=1.3$ mag, when we assume case B  recombination with
$T_{e}=10000\mathrm{K}$ \citep{2006agna.book.....O} and the
extinction law \citep{weingartner01}, while we
get $A_V=3.5$ mag using the recombination line ratio of
Br$\alpha$/H$\alpha$. 
The difference can be
explained by the deeper penetration of Br$\alpha$ photons, suggesting that
only the surface of the emitting region contributes to optical emission
lines.

We calculate the SFR using the PAH 3.3\micron~band, which 
is not expected to be affected by an AGN. With the empirical ratio,
$L_{\mathrm{PAH 3.3}}/L_{IR} \sim 2-3\times 10^{-4}$ for starburst galaxies
\citep{mouri90} and the conversion of the IR luminosity
into SFR \citep{kennicutt98}, the SFR of LEDA 84274 is estimated to be
$20 - 30\
\mathrm{M}_{\odot}\ \mathrm{year}^{-1}$. 
We estimate SFRs of 7 and 25 $\mathrm{M}_{\odot}\ \mathrm{year}^{-1}$
using the optical H$\alpha$ emission line with
$A_{\mathrm{V}}=1.3$ and the NIR Br$\alpha$ emission line with
$A_{\mathrm{V}}=3.5$, respectively. SFR(Br$\alpha$) is consistent with
SFR(PAH), while SFR(H$\alpha$) is lower than the others, which can
be explained similarly by a highly non-uniform $A_{\mathrm{V}}$. 
The infrared luminosity calculated from the IRAS measurement is
$L_{\mathrm{IR}}=1.8\times 10^{11}\ \mathrm{L}_{\sun}$, which
corresponds to SFR(IR)=31 $\mathrm{M}_{\odot}\
\mathrm{year}^{-1}$. 

\begin{figure*}[htbp]
    \resizebox{17cm}{!}{\includegraphics{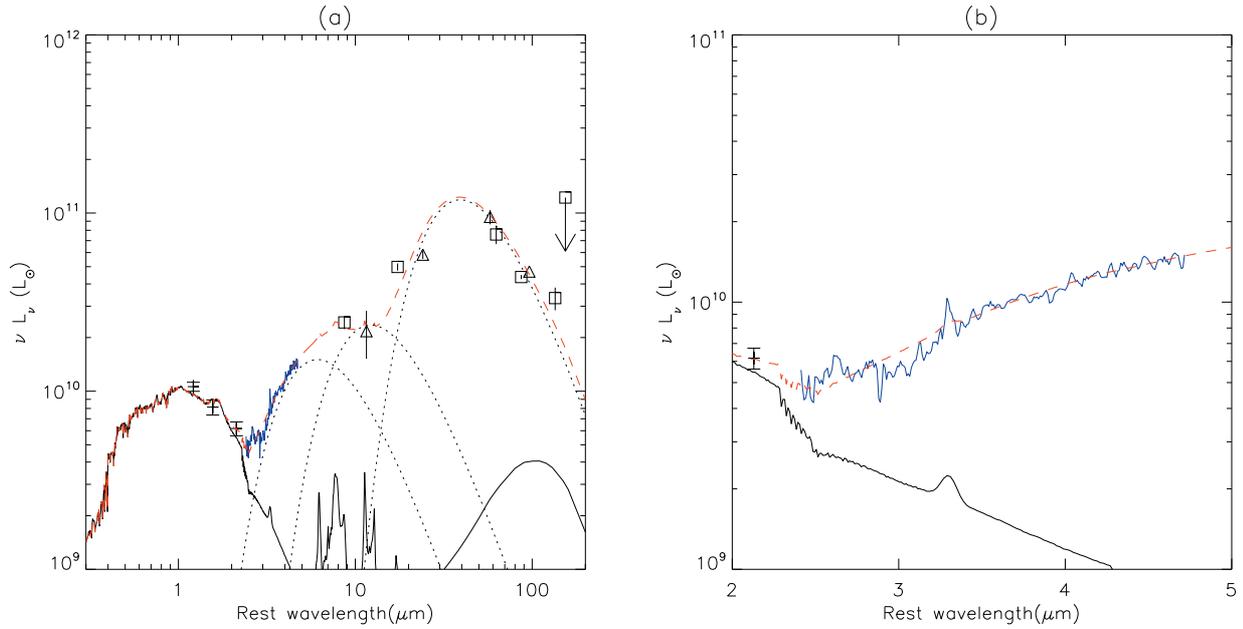}}
    \caption{Spectral energy distribution of (a) LEDA 84274. The blue
      line shows the $AKARI$ NIR spectroscopy. Squares and triangles
      represent the
      $AKARI$ All-Sky Survey and IRAS, respectively. Crosses
      show the 
      2MASS photometry. The solid line(black) is a model
      template of a Sc galaxy \citep{polletta07}. The model line
      normalized at the 2MASS photometry. A dashed line (red) represents the
      Sc galaxy template plus three blackbody components, the
      temperatures of which are 600K, 300K, and 93K. Dotted lines
      represent each blackbody 
      component. (b) Close-up of the 2-5 \micron~range of LEDA 84274,
      to clearly show the model and the blackbody component
      with 600K. See text for details.
    }
    \label{fig:sed}
\end{figure*}

The spectral energy distribution (SED) of LEDA 84274 is presented in
Figure \ref{fig:sed}. A SED model is  also overplotted in Figure
\ref{fig:sed}(a). 
For LEDA 84274, the Sc galaxy template \citep{polletta07}
agrees with the photometric points in
the $J-$, $H-$, and $K_S-$bands. 
To fit the $AKARI$ NIR spectrum, we assume that the 
spectrum is a summation of the Sc galaxy template and dust with a
single-temperature modified blackbody.
The model is defined as
\begin{equation}
    \label{equ:1}
    F_{\lambda}=C_1\ F_{\mathrm{galaxy}}+C_2\ B_{\lambda}(T_{\mathrm{dust}})(1-\exp(-\tau_{\lambda})),
\end{equation}
where $F_{\mathrm{galaxy}}$ is the Sc galaxy template,
$B_{\lambda}(T_{\mathrm{dust}})$ is the Planck function with dust 
temperature $T_{\mathrm{dust}}$ and $\tau_{\lambda}$ is the dust optical
depth. For the frequency dependence of the dust optical depth, we
adopt $\tau_{\lambda} = C_3 \lambda^{-\beta}$.  
The parameters $C_1, C_2, T_{\mathrm{dust}}, C_3$,
and $\beta$
are derived by $\chi^2$ fitting the $AKARI$ NIR spectrum. 
The fit indicates that $530\mathrm{K}<T_{\mathrm{dust}}<670\mathrm{K}$
if optically thick, or
$470\mathrm{K}<T_{\mathrm{dust}}<580\mathrm{K}$ if optically thin
with $\tau_{1\mu\mathrm{m}} \lesssim 1$ and $\beta=1-2$. 
Figure \ref{fig:sed} (b) shows a close-up of a fit with
an optically thick blackbody of $T_{\mathrm{dust}}=600K$.

In addition, we must add a cool component of dust with
$T_{\mathrm{dust}}=93\pm5\ \mathrm{K}$ and $\beta=0$ to
explain the IRAS and $AKARI$ photometric data at wavelengths longer than 12\micron. For $AKARI$ 9\micron~and IRAS 12\micron~data, we need to
add one more component of temperature 
$T_{\mathrm{dust}}\sim 300\mathrm{K}$.
There is no strong
constraint on either the temperature or the emissivity. 

The fitting results are shown in Figure \ref{fig:sed} (a). From the
blackbody fits, we derived a total infrared luminosity of $L_{\mathrm{IR}}=1.8
\times 10^{11} \mathrm{L}_{\sun}$ between 8 and 1000
\micron~, which is consistent with the $L_{\mathrm{IR}}$ derived using
the method of \citet{kim95}. We note that the total luminosity is dominated by the cool
component.  


\subsection{IRAS 01250+2832}
\label{sec:iras}

\begin{figure*}[htbp]
    \resizebox{17cm}{!}{\includegraphics{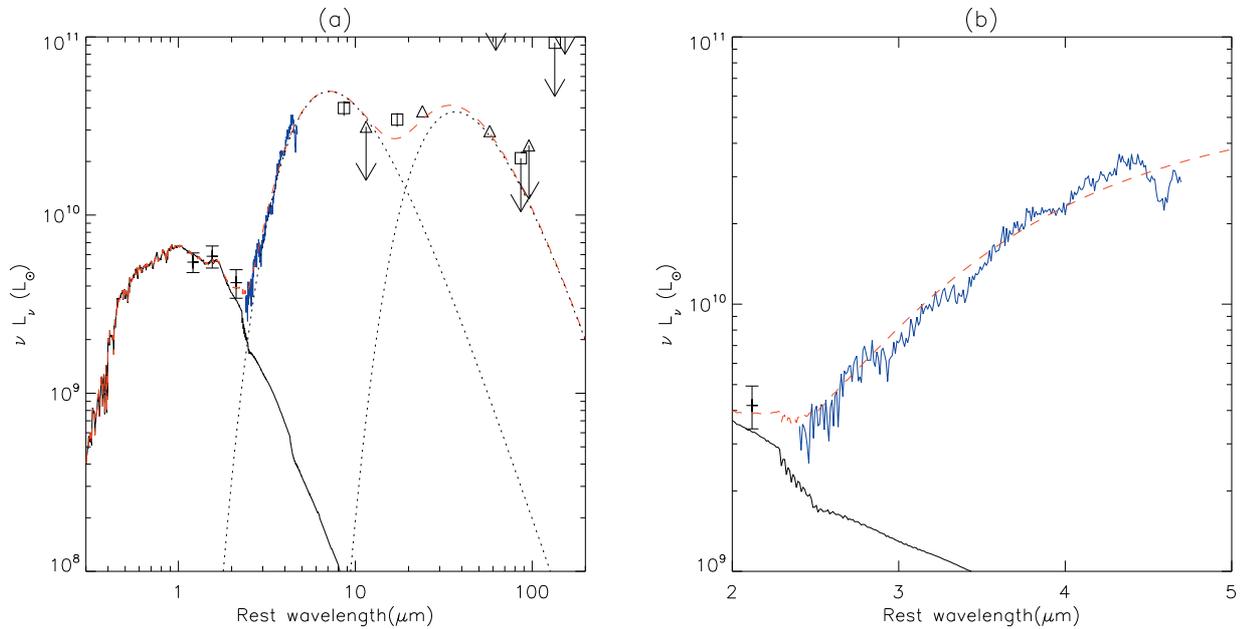}}
    \caption{Same as Figure \ref{fig:sed2} except for IRAS
      01250+2832. A model
      template of an elliptical galaxy \citep{silva98} is used. Two
      blackbodies in this figure have temperatures of 510K and 100K.
    }
    \label{fig:sed2}
\end{figure*}

The SED of IRAS 01250+2832 is shown in
Figure \ref{fig:sed2}.
It is difficult to explain the flat SED at
wavelengths between 
9\micron~and 60\micron~with normal spiral and
elliptical 
galaxy models\citep{silva98}. 
The spiral and elliptical galaxy
templates do not have strong enough 
infrared emission. 
It is difficult to explain the global SED
with starburst galaxy templates such as 
M82 and Arp 220 that have a strong excess in the
FIR. Here, we assume that the host galaxy is 
dominated by an
old stellar population
\citep[13Gyr;][]{silva98} based on the strong
$D_{\mathrm{n}}(4000)$ index. 

When we also fit the $AKARI$ NIR
spectrum using Equation \ref{equ:1} and the elliptical galaxy template, 
we need to introduce a blackbody
component of temperature of $510\pm20 \mathrm{K}$ in 
optically thick conditions as shown in Figure \ref{fig:sed2} (b). 
Another blackbody component of $100\mathrm{K}$ is needed
to explain the $IRAS$ 60\micron~flux.
The higher fluxes of the model relative to those observed at 9 and 12\,$\mu$m can be
attributed to
9.7\micron~silicate absorption,
which is not included in this model. However, it is expected to be present
based on the CO absorption measured at 4.75\micron.
The conclusion that the SED needs two blackbody
components does not change much even if we use spiral galaxy templates. 
The infrared luminosity $L_{\mathrm{IR}}$ is derived to be 
$9.1\times 10^{10}\ \mathrm{L}_{\sun}$, which is consistent with
$1.0 \times 10^{11}\ \mathrm{L}_{\sun}$ using the method of
\citet{kim95}. We note that the contribution at wavelengths longer
than 60\micron~to the total luminosity is expected to be small for
IRAS 01250+2832.  


While there are no signs of H$\beta$ and [OIII] emission lines in
Figure \ref{fig:opt_spec}(b), 
H$\alpha$ and [NII] lines are clearly detected. The [NII]6584\AA~
emission line is stronger than H$\alpha$, suggesting that IRAS
01250+2832 is a low-ionization nuclear emission-line region (LINER)
galaxy. However, the other indicators of LINERs, [OI]6300\AA~and
[SII]6716\AA+6731\AA~emission lines, are not seen because of the
weakness of the lines, relative to the stellar continuum. 
The 3$\sigma$ upper limit to the
H$\beta$ flux is estimated to be 3.0$\times 10^{-16}\
\mathrm{ergs}\ \mathrm{s}^{-1}\ \mathrm{cm}^{-2}$. This upper limit
is not useful to estimate the reddening from the Balmer
ratio. Without applying any reddening correction, we calculate an H$\alpha$
line luminosity of 4.7$\times 10^5\ \mathrm{L}_{\sun}$, which is
comparable to that of low-luminosity AGNs (LLAGNs) found in the
optical spectroscopic survey of nearby
galaxies \citep{ho97}. A comparison with LLAGNs is discussed in
Section \ref{sec:dis}.

\section{Hot dust with $T_{\mathrm{dust}}\sim 500-600\mathrm{K}$}
\label{sec:hot}

As shown in Section \ref{sec:multi}, both galaxies show near-infrared continuum
from hot dust with $T_{\mathrm{dust}}=500-600\mathrm{K}$, 
although they differ in other ways. 
The origin of the energy source heating this dust, however, remains unclear.

Fits
indicate that the luminosity of this hot dust component
is $2.7 \times 10^{10}\ \mathrm{L}_{\odot}$
and $6.6 \times 10^{10}\ \mathrm{L}_{\odot}$ for LEDA 84274 and IRAS
01250+2832, respectively. 
If we assume 
one single optically thick spherical emitting region, the size of the emitting
regions would be
0.7 pc and 1.6 pc in diameter for LEDA 84274 and IRAS 01250+2832,
respectively. 
One possibility is that we are observing an OB star cluster that has a huge luminosity
of $L_{\mathrm{IR}} > 10^{10} \mathrm{L}_{\odot}$ in a 
compact volume of a cubic parsec, and that heats up the
surrounding dust to $>$500K.  
However, the relaxation time of the dynamically bound system can be
estimated to be
$10^6 - 10^7$ years. Thus, this situation is unlikely to be observed.

A more likely possibility is that the object is a compact central supermassive black hole 
surrounded by a hot accretion disk.
However, no optical spectrum shows any evidence of the
hard photons produced by
an AGN. The various emission-line ratios of LEDA 84274 in the optical band show that it is an
HII-region-like galaxy as 
\citet{kim95} concluded. IRAS 01250+2832 shows no emission lines
except for H$\alpha$ and [NII]. 
The weak- and non-detection of PAH 3.3\micron~emission may support the
presence of an obscured AGN. 
The equivalent widths of the PAH 3.3 \micron~emission feature are
0.018 \micron\,
and $<0.003$ \micron~for LEDA 84274 and IRAS 01250+2832, respectively. 
Observations of starbursts show equivalent widths of PAH
3.3 \micron~with an
average value of 0.1\micron, and they are never smaller than 0.04\micron
\citep{moorwood86,imanishi08}.   Thus the PAH equivalent width
is diluted by strong featureless NIR continuum in IRAS 01250+2832
and completely overwhelmed in LEDA 84274.


The excess featureless near-IR continuum
may alternatively represent multiple dusty OB star
clusters, in which extraordinarily intense starbursts are occurring. This is seen
in some interacting
galaxies\citep{joseph84,joseph85}. \citet{hunt02} also
argue that the hot dust that 
produces the NIR color excess must be heated by massive stars. In this
case, the weak PAH emission can be produced by the harder
radiation fields of massive stars in these environments that can destroy
PAHs \citep[e.g.][]{beiro06,wu06}. From $Spitzer$ imaging data of
nearby irregular galaxies, 
excess emission at 4.5 \micron~from hot dust was found
mostly in high-surface-brightness HII regions, implying that massive stars are the
primary source of heating \citep{hunter06}. 
For IRAS 01250+2832, multiple dusty star clusters are less
likely to be present, because most of
dusty star clusters are surrounded by diffuse emission of PAHs as
well as optical emission lines.  
IRAS 01250+2832 does not show any evidence for such activities.

If the assumption that the dust emission is optically thick is not
correct, 
the NIR red continua of LEDA 84274
can be explained by a dust temperature of $\sim 480-580$K and
emissivity of $\lambda^{-\beta}$ with $\beta=1-2$.
In spiral galaxies that normally display the thermal emission of optically
thin dust, \citet{lu03} detected non-stellar NIR excess 
continuum with a temperature of $\sim
1000$K($\beta=2$)
and concluded that the NIR excess continuum originates in the interstellar
matter of the galaxies based on the linear correlation between emission
from aromatic carbon and the excess. 
In this case, the NIR excess has a luminosity of only a few percent
of the FIR luminosity.
However, the ratio of the NIR luminosity to the FIR luminosity is
0.15 and 0.75 for LEDA 84274 and IRAS 01250+2832, respectively.  
Thus, the emission from the hot dust in LEDA 84274 and IRAS 01250+2832
seems to be different from those of spiral galaxies.

We suspect that the hot dust associated with obscured
AGNs. But 
the possibility of multiple dusty star clusters in both objects, especially LEDA 84274,
cannot be ruled out. 
Observations at other wavelengths can resolve the question.
In particular, X-ray observations can provide important
information on the energy source, but both
galaxies are not detected in the ROSAT
all-sky survey faint source catalogue \citep[an upper limit of $1.3
\times 10^{-13}\ \mathrm{ergs}\ \mathrm{cm}^{-2}$ in the 0.1-2.4 keV
band;][]{voges00}.  Hard X-ray observations will thus be important.
 
\section{Discussion}
\label{sec:dis}

As discussed in Section \ref{sec:hot}, the compact hot dust component
that we found in LEDA 
84274 and IRAS 01250+2832 is likely to surround
an AGN that is heating it, even though other possibilities cannot be completely ruled
out. In this section, we 
discuss these dusty AGNs. 

The optical spectra display no strong emission lines (especially
high-excitation ones) from narrow-line regions. 
Thus, no narrow-line region on a 10-1000pc scale is evident in
these galaxies. 
IRAS 01250+2832 has a low-ionization
spectrum that is typical of  
LLAGNs, as discussed in Section \ref{sec:iras}. The low-ionization
state spectrum in LLAGNs may be produced by black-hole accretion at a very low
rate (with a bolometric luminosity
of $L_{bolo}<4\times 10^{10}\ \mathrm{L}_{\sun}$\citep{ho08}). However, LEDA 84274
and IRAS 01250+2832 have much higher luminosities of $L_{\mathrm{IR}}
\gtrsim 1 \times 10^{11}\ \mathrm{L}_{\sun}$ at infrared
wavelengths than LLAGNs. 
Large optical extinction of the
central engine is a plausible explanation of
their optical spectra. 
These findings indicate that the AGNs are surrounded by a large amount of
dust, so that ionizing UV radiation from the AGN is blocked along virtually 
all lines of sight. In ultra-luminous and luminous infrared galaxy
cores, similar situations are sometimes seen \citep[e.g.][]{imanishi10}.

Since the observed radiation from AGNs is often reprocessed and
re-emitted at infrared wavelengths, the intrinsic
SED is altered substantially. Wide spectral coverage is necessary to determine the
Eddington ratio ($L_{\mathrm{bolo}}/L_{\mathrm{Edd}}$, or equivalently
$L_{\mathrm{bolo}}/M_{\mathrm{BH}}$) \citep[e.g.][]{peterson97}. 
Although it is difficult to differentiate the
AGN component from the overall SED, we estimate the black-hole mass by
making the
a simple assumption that the AGN luminosity is nearly equal to the
luminosity from the 500-600K dust.
Assuming spherical accretion toward the black hole with a
radiative efficiency of 0.1,
the supermassive black hole mass from the luminosities of the hot
dust is at least  $8\times 10^6\ 
\mathrm{M}_{\odot}$  and $2\times 10^7\ \mathrm{M}_{\odot}$ for LEDA
84274 and IRAS 01250+2832, respectively. 
These do not differ significantly from the black hole masses
estimated 
from the 9\micron~luminosities using the bolometric correction
evaluated using
a Type2 QSO template in \citet{polletta07}
\footnote{The
  hot dust luminosities estimated from 
  the 500-600K blackbody are 30 percent
  higher and 4 percent lower than those estimated from the bolometric
  correction, for LEDA 84274 and IRAS
  01250+2832, respectively.}. Thus, we adopt the black hole mass
estimates from the 
luminosities of the hot dust in the following discussion.
If they have a higher radiative
efficiency, the masses will be lower. 

The masses of these host
galaxies can be calculated to be $6\times 10^9\ \mathrm{M}_{\odot}$ and $4
\times 10^{9}\ \mathrm{M}_{\odot}$  for LEDA 84274 and IRAS
01250+2832, respectively, using the 2MASS {$K_S$} photometry and the
mass-to-luminosity ratio, $M_{\mathrm{host}}/L_{K_S} \sim 1$
\citep{cole01}\footnote{In \citet{cole01}, mean stellar mass-to-light
  ratios are estimated to be 0.73 and 1.32
  $\mathrm{M}_{\sun}/\mathrm{L}_{\sun}$ for the Kennicutt and the
  Salpeter initial mass functions, respectively.}. The
mass ratio of the black hole to the stellar component in the host
galaxy is $1.3 \times 10^{-3}$ and $5 \times 10 ^{-3}$ for LEDA 84274
and IRAS 
01250+2832. The results are consistent with the local relation between
the mass of the central black hole and the stellar mass of the
surrounding spheroid or the bulge in nearby galaxies,
$M_{bh}/M_{bulge}=1.4 \times 10^{-3}$ \citep{haring04}.
The mass of both host
galaxies is relatively low, i.e., $6\times 10^9\ \mathrm{M}_{\odot}$ and $4
\times 10^{9}\ \mathrm{M}_{\odot}$,  for LEDA 84274 and IRAS
01250+2832, respectively. 
Large samples of AGNs from the Sloan Digital Sky Survey have
host galaxy masses of $\log M_{gal}=9.5-12$ with $D_\mathrm{n}(4000)=
1.2-2.2$ \citep{kauffmann03}. Our sample with $M_{\mathrm{host}}=4-6 \times 10^9\
\mathrm{M}_{\odot}$ and $D_n(4000) = 1.1 - 1.6$ seems to be the least
massive population that may harbor an AGN. 


\section{Summary}
\label{sec:sum}

By combining data from the $AKARI$ MIR All-Sky Survey and NIR
spectroscopy, we have discovered two luminous infrared
galaxies at 
z$\sim 0.04$ that have hot dust with a temperature of $\gtrsim$500K. 
The hot dust is likely to be
associated with AGNs, but the possibility that the hot dust is heated by multiple dusty star clusters of massive
stars cannot be ruled out by our current data. If they are AGNs,
their emission must  be buried, because present AGN
signatures are seen  
only in the NIR and MIR, not in the optical. 
If they have AGNs, the estimates of the black hole masses of
$\sim 8-20 \times 10^7\ \mathrm{M_{\odot}}$, and
the host galaxy masses of $\sim 4-6 \times 10^9\ \mathrm{M_{\odot}}$ 
also indicate that the population is less massive than optically selected AGNs
\citep{kauffmann03}.  
These objects were missed in previous surveys, demonstrating the
power of $AKARI$ MIR All-Sky Surveys to obtain a more complete view of
the entire AGN population. 

We have initiated a long-standing program to study AGNs detected from the
$AKARI$ MIR All-Sky Survey. Multiwavelength observations
will provide larger
samples of buried AGN candidates that are similar to those we present in this
paper, examples of which we plan to present in the near future. 

\begin{acknowledgements}
The authors thank the anonymous referee for constructive comments that
helped to improve this paper. 
This research is based on observations with $AKARI$, a JAXA project with
the participation of ESA.
This research also makes use of data products from the Two Micron All
Sky Survey, which is a joint project of the University of
Massachusetts and the Infrared Processing and Analysis
Center/California Institute of Technology, funded by the National
Aeronautics and Space Administration and the National Science
Foundation. SO has been supported by the a Grand-in-Aid for Scientific
Research from the Japan Society for the Promotion of Science
(No. 10018958). DI is supported by the Nagoya University Global COE
Program ``Quest for Fundamental Principles in the Universe'' from the
Ministry of Education, Culture, Sports, Science, and Technology of Japan.
\end{acknowledgements}

\end{document}